\documentclass{sf2a-conf2013}
\usepackage{graphicx}
\usepackage{hyperref}
\usepackage[]{natbib}  
\usepackage{epstopdf}
\usepackage[]{units}

\newcommand{\hess}{\textsc{H.E.S.S.}}
\newcommand{\magic}{\textsc{MAGIC}}
\newcommand{\veritas}{\textsc{VERITAS}}
\newcommand{\fermi}{\textsc{{\it Fermi}-LAT}}

\def\BibTeX{{\rm B\kern-.05em{\sc i\kern-.025em b}\kern-.08em
    T\kern-.1667em\lower.7ex\hbox{E}\kern-.125emX}}
\bibpunct{(}{)}{;}{a}{}{,}  

\begin{document}

\TitreGlobal{SF2A 2013}


\title{Constraining gamma-ray propagation on cosmic distances}

\runningtitle{Cosmic gamma-ray propagation}

\author{J. Biteau$^{1,}$}\address{Santa Cruz Institute for Particle Physics, Department of Physics, University of California at Santa Cruz, Santa
Cruz, CA 95064, USA}\address{Laboratoire Leprince-Ringuet, Ecole polytechnique, CNRS/IN2P3, F-91128 Palaiseau, France}

\setcounter{page}{237}


\maketitle

\begin{abstract}
Studying the propagation of gamma rays on cosmological distances encompasses a variety of scientific fields, focusing on diffuse radiation fields such as the extragalactic background light, on the probe of the magnetism of the Universe on large scales, and on physics beyond the standard models of cosmology and particle physics. The measurements, constraints and hints from observations of gamma-ray blazars by airborne and ground-based instruments are briefly reviewed. These observations point to gamma-ray cosmology as one of the major science cases of the Cherenkov Telescope Array, CTA.
\end{abstract}


Gamma-ray astronomy at very high energies (VHE, $\unit[30]{GeV}<E<\unit[30]{TeV}$) is driven by ground-based imaging atmospheric Cherenkov telescopes, whose current major representatives are \hess\ (Namibia), \magic\ (Canaries islands) and \veritas\ (Arizona). Together with airborne instruments observing the gamma-ray sky at high energies (HE, $\unit[300]{MeV}<E<\unit[300]{GeV}$), such as \fermi, they can now probe and constrain the propagation on billion-lightyear distances of the highest-energy photons produced and detected in the Universe. The observational field called gamma-ray cosmology encompasses the study of the extragalactic background light (EBL), an optical-infrared radiative component born with the first stars and that kept on growing since then, of the intergalactic magnetic field, which fills the voids between large scale structures and whose origin remains unidentified, and of physics beyond the standard models of cosmology and particle physics, possibly revealing hints on  dark matter candidates and on the laws of nature at the Planck scale. 

The science case of gamma-ray cosmology is discussed in Sec.~\ref{Sec:ScienceCase}. A wide-spectrum but non-exhaustive list of current constraints, hints and measurements in the field is exposed in Sec.~\ref{Sec:Constraints}. Possible refinements in the analyses are discussed in Sec.~\ref{Sec:Future}, together with the great expectations from the next generation instrument, the Cherenkov Telescope Array (CTA).

\section{Science case of gamma-ray cosmology}\label{Sec:ScienceCase}

Gamma-ray cosmology could be considered as an established domain, with seminal theoretical studies that can be traced back to the '60s \citep{REF::NIKISHOV::JETP1962,1966PhRvL..16..479J,1967PhRv..155.1408G}, but the first powerful enough extragalactic 
accelerators to be used as probes were detected in the beginning of the '90s. These strong sources are discussed in the following subsection together with the science case addressed. 

\subsection{Extragalactic sources as gamma-ray beacons}

\begin{figure}[ht!]
 \centering
 \includegraphics[width=0.49\textwidth,clip]{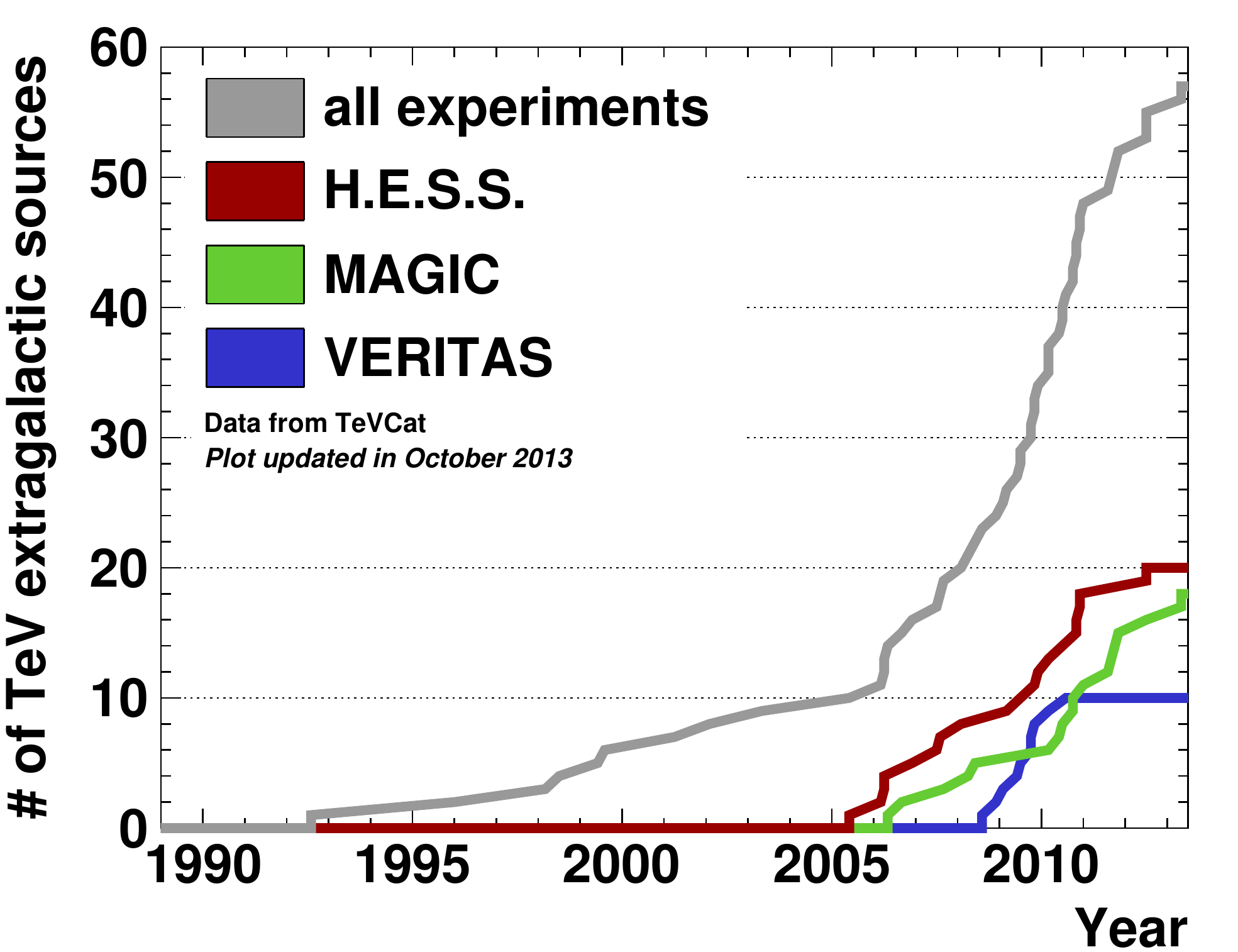}%
 \hfill \includegraphics[width=0.49\textwidth,clip]{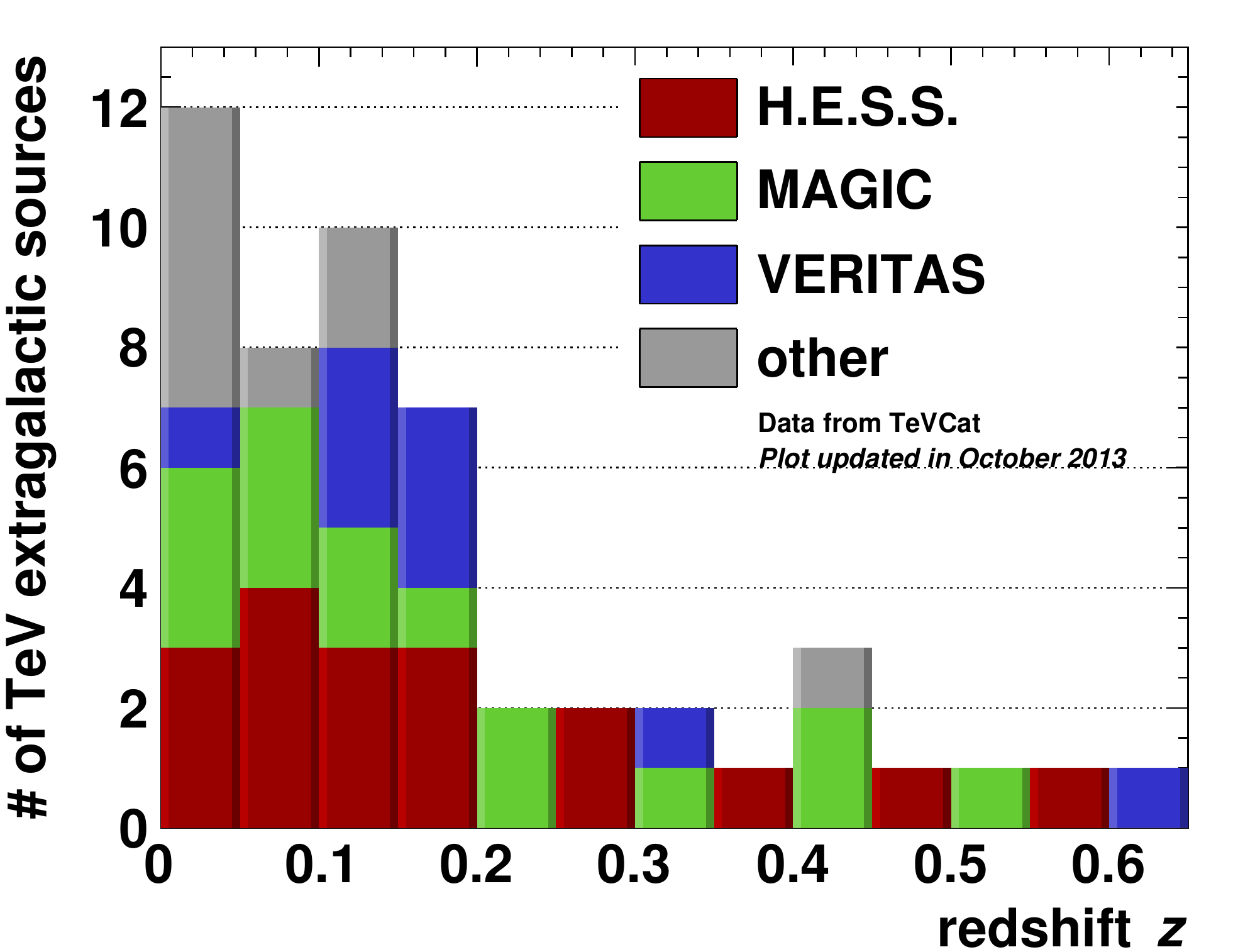}      
  \caption{{\bf Left:} Number of detected extragalactic emitter of VHE gamma rays as a function of the year of discovery. {\bf Right:} Distribution of the redshifts of the current population of extragalactic VHE emitters with a constrained distance. {\it Data from TeVCat}\ \url{http://tevcat.in2p3.fr}.}
  \label{Biteau:fig2}
\end{figure}

Since the groundbreaking detection of Mrk~421 \citep{Punch92} by the first generation ground-based instruments Whipple, the number of 
detected VHE extragalactic sources has exponentially grown, up to almost 60 sources in mid 2013. CAT and HEGRA, the major
second generation instruments, contributed to the effort, yielding a handful of sources in the beginning of the 2000s. As
shown in Fig.~\ref{Biteau:fig2}, left, the third generation instruments \hess, \magic\ and \veritas\ have revolutionized the field
with the discovery of almost 50 sources in less than ten years, up to redshifts $z>0.6$, as shown in Fig.~\ref{Biteau:fig2}, right.

Besides the nearby starburst galaxies NGC~253 and M~82, the VHE extragalactic population is exclusively composed of active galactic nuclei (AGN). These objects are thought to host a supermassive black hole, possibly as heavy as a billion solar masses, fed by an accretion disk, with the emission of relativistic jets on each side of the disk for the brightest radio-emitters \citep[see][for a unified picture of AGN]{1993ARA&A..31..473A,1995PASP..107..803U}. With a typical opening angle of few degrees, these jets, which are composed of magnetic fields, accelerated particles and beamed radiation, can either be seen edge on, as for the nearby radio galaxies M~87, Centaurus~A and NGC~1275, or they can be closely aligned with the line of sight. In the latter case, the AGN is called a blazar and it appears particularly bright and energetic due to the relativistic beaming and the Doppler shift of the energy of the photons \cite[see e.g.][for a recent discussion on gamma-ray blazars]{2013arXiv1309.4772G}.

Blazars constitute the vast majority of detected AGN at VHE and can be divided in flat-spectrum radio quasars (3 detected at VHE), which exhibit strong emission lines, and BL Lac objects (49 detected at VHE), with weak or absent lines. The broad band emission of blazars is composed of a radio to UV/X-ray component, generally attributed to the synchrotron emission of relativistic electrons in the jet, and of a gamma-ray component, whose origin remains debated. The simplest radiative models attribute it to the Comptonization of the synchrotron field by the very electrons that generated it. This synchrotron self-Compton model \citep{1985ApJ...298..128B} works fairly well for the largest subclass of BL Lac objects, the high-synchrotron-frequency-peaked objects (HSP, sometimes HBL in the literature), whose low-energy components peak in X rays. For objects with a lower-energy synchrotron peak, including flat-spectrum radio quasars, external photon fields coming from the accretion disk (among other regions) could be scattered by the electrons resulting in additional radiative components. The potential correlation of these fields with the black-hole and accretion-rate evolution reduces the value of low-frequency-peaked objects as reliable cosmological gamma-ray beacons \citep{Reimer2007}. Instead, the apparent simplicity of the emission of HSP blazars (with the caveat of their flux variability) makes them the best reference emitters so far to constrain gamma-ray propagation.

\subsection{Pair creation on the extragalactic background light}

The leading effect on gamma-ray propagation is pair creation on diffuse radiation fields in the Universe. For an isotropic target field of fixed energy $\epsilon$, the cross section is maximum in the comoving frame for a gamma-ray energy $E \sim (2m_ec^2)^2/\epsilon \sim \unit[1]{TeV} \times (\epsilon/\unit[1]{eV})^{-1}$ \citep{1976tper.book.....J}. TeV gamma rays thus strongly interact with eV, i.e. micron-wavelength photons. This is precisely the wavelength where the cosmic optical background is most intense. The cosmic optical background extends from UV to near infrared wavelengths and is composed of the total emission of stars and galaxies since the end of the cosmic dark ages. Its counterpart at lower energy is the cosmic infrared background, with a peak intensity around $\sim\unit[100]{\mu m}$, which comes from the reprocessing of UV-optical light by dust. The cosmic optical and infrared backgrounds are usually referred to as the extragalactic background light (EBL) and constitute the second most intense cosmological background after the CMB, which peaks at even lower energy.

Measuring the EBL directly proves to be difficult because of the foregrounds from the Galaxy and the solar system (zodiacal light) but stringent constraints can be derived accumulating the brightness of known galaxies \citep[e.g.,][]{Dole,2000MNRAS.312L...9M,2004ApJS..154...39F}. Lower limits from galaxy counts typically lie one order of magnitude below upper limits derived from direct measurements. The lower limits are reproduced by EBL models based, e.g., on the local population of galaxies \citep{Fr08}, on the evolution of large samples \citep{Dominguez},
or on semi-analytical models \citep{Gilmore}. Comprehensive summaries of the observational techniques at these wavelengths as well as of the modeling approaches can be found in \citet{HauserDwek}, or in \cite{2013APh....43..112D} for a more recent review.

To date, the best way to make local foregrounds negligible consists in integrating the optical/near-infrared emission over cosmological distances, as made possible by gamma-ray observations. Indeed, for a density of photons $n(\epsilon,z')$ in the energy band $[\epsilon;\epsilon+{\rm d}\epsilon]$ and at a redshift $z'$ (i.e. $\int{\rm d}\epsilon\ n(\epsilon,z') $ is a number of photons per unit volume), gamma rays with observed energy $E$, emitted by a source at a redshift $z$, are absorbed (through pair creation) with an optical depth:
\begin{equation}
\tau\left(E,z\right) = \int_0^z{\rm d}z' {{{\rm d}l} \over {{\rm d}z}}(z') \int_{0}^{+\infty}{\rm d}\epsilon\ n(\epsilon,z') \int_{-1}^{1}{\rm d}\mu {{1-\mu} \over 2} \sigma_{ee}(\epsilon,E\times(1+z'),\mu)
\label{Eq:EBLOptDep}
\end{equation}

\begin{figure}[ht!]
 \centering
 \includegraphics[width=0.65\textwidth,clip]{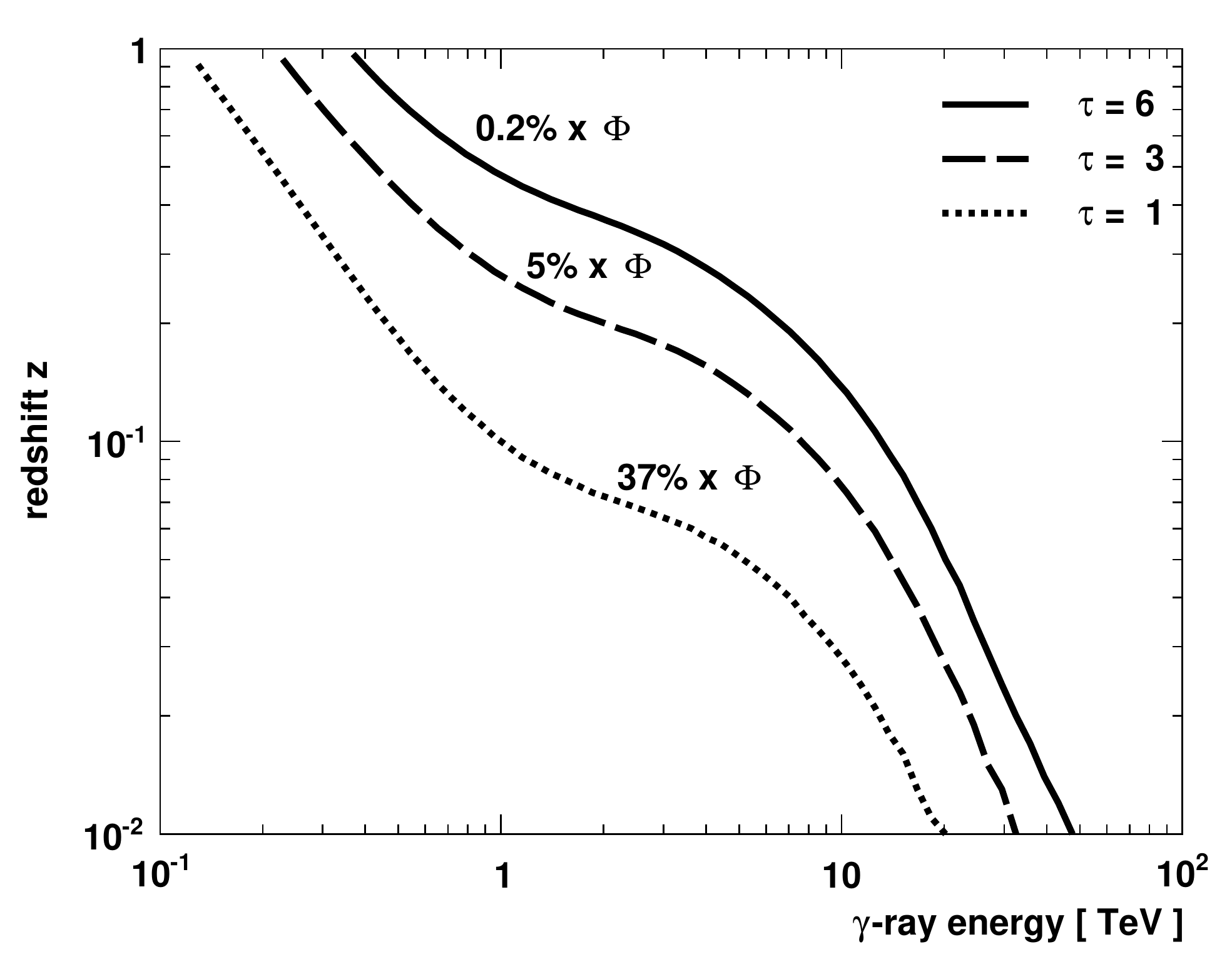}%
  \caption{Optical depth $\tau = 1, 3, 6$ as a function of redshift and observed gamma-ray energy, within the modeling of \cite{Fr08}. The frontiers correspond to an observed flux equal to 37\%, 5\% and 0.2\% of the emitted flux.}
  \label{Biteau:fig2b}
\end{figure}

The first integral represents the distance, with ${{{\rm d}l} / {{\rm d}z}} = { c / {H_0(1+z)}} { {\sqrt{\Omega_\Lambda + \Omega_m(1+z)^3} } }$ in a flat universe. The second is the target density and the third is the cross section with an integration in the center-of-mass frame over the angle $\theta$ ($\mu = \cos \theta$) between the directions of the target photon and of the gamma ray. The pair-creation cross section is given by the Bethe-Heitler formula:
\begin{equation}
\sigma_{ee}(\epsilon_1,\epsilon_2,\mu) = {3 \over 16} \sigma_{\rm T} (1-\beta^2)\left[2\beta(\beta^2-2)+(3-\beta^4)\ln\left({{1+\beta}\over{1-\beta}}\right)\right] \Theta(\epsilon-\epsilon_{\rm th})
\end{equation}
where $\Theta$ is the Heaviside function, $\epsilon_{\rm th} = 2m_e^2c^4 / (1-\mu)\epsilon_2 $ is the threshold energy and $\beta^2 =  1 - \epsilon_{\rm th} /  \epsilon_1$.  The pair creation on the EBL thus results in an energy and redshift dependent absorption shown in Fig.~\ref{Biteau:fig2b} that can be used to probe the EBL itself (see Sec.~\ref{Sec:EBLmeasurements}).

\subsection{Intergalactic magnetic field and the reprocessing of the pairs}\label{Sec:IGMFth}

A second order effect on gamma-ray propagation is related to the fate of the pairs created by EBL absorption. 
For a gamma ray of energy $E\sim\unit[1]{TeV}$, each created lepton carries an energy of half a TeV, corresponding to a Lorentz factor $\gamma \sim 10^6$, and can upscatter the most intense photon field in its environment, that is the CMB with photons of energy $\epsilon_{\rm CMB} \sim \unit[1]{meV}$. In the Thompson regime, the resulting photons carry an energy $E'\sim\gamma^2\epsilon_{\rm CMB}\sim \unit[1]{GeV}$. Thus there should be a broad band reprocessing of the signal detected by ground-based VHE instruments into the HE band observed by airborne instruments\footnote{Plasma instabilities resulting in a heating of the medium have nonetheless been invoked as an alternative for the leptons to lose their energy \citep{2012ApJ...752...22B}. The theoretical debate remains open to date.}, an effect often referred to as pair echo. If the intergalactic magnetic field (IGMF) filling the voids is strong enough (typically between $10^{-15}-\unit[10^{-12}]{G}$), the pairs should be significantly deflected resulting in an extended emission around the point-like blazar, an emission sometimes referred to as pair halo or beam broadened cascade.

The intensity and coherence scale of the IGMF remain largely unknown, though these quantities could be of cosmological importance \citep[see][for a recent review of the observational and theoretical progress]{2013A&ARv..21...62D}. To be amplified through dynamo and compression effects up to their current level, the magnetic fields found in galaxies and galaxy clusters would indeed need seeds. These magnetic seeds could arise either during collapses forming the first structures or during the early Universe, be it inflation or electroweak/QCD transition phases. Current constraints from gamma-ray astronomy are discussed in Sec~\ref{Sec:IGMFconstraints}.

\subsection{Alternatives beyond the Standard Models}\label{Sec:BSMth}

Exotic physics beyond the standard models of particle physics and cosmology can also be probed within gamma-ray cosmology. Lorentz invariance violation, studied e.g. within quantum gravity frameworks, would result in an alteration of the dispersion relation for VHE gamma rays. The threshold of the pair creation, derived with simple special relativity arguments, would thus be altered, reducing the EBL opacity to gamma rays \citep{1999ApJ...518L..21K,2008PhRvD..78l4010J}. The absence of such a feature in the spectrum of Mrk~501, as detected by second generation instruments, already set interesting limits \citep{2001APh....16...97S} but the way remains wide open for refinements with current data.

Another crucial test concerns axion-like particles (ALPs). These belong to the class of weakly interacting slim particles (WISPs) that remain a viable alternative to their massive counterparts (WIMPs) as elementary constituents of cold dark matter \citep[see e.g.][in the context of the future instrument ALPS-II]{2013JInst...8.9001B}. Photons and ALPs would indeed mix in the presence of a magnetic field (such as the IGMF or intracluster magnetic fields), which could modify the apparent absorption of gamma rays \citep[e.g.,][]{2009PhRvD..79l3511S}. A gamma ray converted in an ALP would be unaffected by the EBL absorption and could propagate on large distances before converting back into a gamma ray and being detected. The universe would then appear more transparent than expected. Alternatively, for low enough reconversion probabilities, the ALPs would go undetected, potentially increasing the opacity to gamma rays. Recent hints of deviations from a standard EBL absorption (so-called pair production anomaly) are currently debated in the community, as discussed in Sec.~\ref{Sec:BSMhints}.

\section{First measurements, hints and constraints}\label{Sec:Constraints}

\subsection{A short history of gamma-ray cosmology}\label{Sec:History}

The methodology to constrain EBL with VHE gamma-ray blazars was developed with the flat spectrum radio quasar 3C~279, which would eventually be detected with the sensitivity of third generation instruments. \cite{1992ApJ...390L..49S} performed a straight extrapolation of the HE spectrum of 3C~279 as measured by EGRET up to VHE, arguing that the object should be detectable by the Whipple observatory under the assumption of no curvature and no EBL absorption. The non detection proved either an intrinsic curvature in the spectrum or/and that EBL absorption, which increases with energy (cf Fig.~\ref{Biteau:fig2b}), is strong enough to suppress the flux. A characterization of the flux suppression could then provide constraints on the EBL.

The detection of Mrk~421 by Whipple enabled the first derivations of upper limits on the EBL \citep{1993AAS...183.2906S,1994ApJ...436..696D,1998PhRvL..80.2992B}. Further studies followed the detection of Mrk~501 by Whipple and HEGRA \citep{1998APh.....9...97F,1999A&A...349...11A}, and later on by CAT \citep{2000A&A...359..419G,2001A&A...371..771R}, then complemented by data from the more distant ($z\sim0.13$) HSP blazar H~1426+428 \citep{Dwek}. More stringent constraints came from third generation instruments, in particular with the detection of 1ES~1101-232 by \hess\ \citep{EBLAHA} and the combination of spectra from 13 blazars by \cite{EBLMaRa}. Absorption at the low energy end of the VHE band has also been constrained by \fermi\ using high redshift blazars and GRBs \citep{UpperLimitFermi}.

The combination of HE and VHE measurements led to the most constraining upper limits to date. For $z<0.5$, the HE band is indeed virtually unaffected by EBL absorption, and \fermi\ data can be used to constrain the intrinsic emission, while VHE data probe the absorption \citep[as, e.g., in][]{Georganopoulos,Orr,Meyer12}. 

\subsection{Measurements of the EBL}\label{Sec:EBLmeasurements}

One of the limitations of the studies mentioned above lies in the method originally developed for 3C~279. Since broad band EBL absorption and the intrinsic curvature expected in standard emission scenarios both result in downward going observed flux as a function of energy, a characterization of the fall can only result in an upper limit on the EBL absorption, derived on the basis of a null intrinsic curvature.

This limitation has been tackled by the \fermi\ and \hess\ collaborations with two slightly different approaches. In \cite{2012Sci...338.1190A}, the curvature in the spectra of BL Lac objects was measured in an energy range where the EBL is ineffective (typically $\unit[1-50]{GeV}$ for $0.5<z<1.6$), modeling it with the simplest functional form: a log parabola, i.e. a parabola in log-log scale. Extrapolating the curvature up to $\unit[500]{GeV}$ enables the derivation of the remaining absorption. To do so, a maximum likelihood approach is adopted, each spectrum being fitted with a model:

\begin{equation}
\phi(E,z) = \phi_{\rm int}(E) \times \exp(-\alpha\tau(E,z,n))
\label{Eq:SpecModel}
\end{equation}
where $\phi$ is the measured flux, $\phi_{\rm int}$ is the extrapolated curved spectrum, the so-called intrinsic spectrum, $\tau(E,z,n)$ is the EBL optical depth for a template EBL model of density $n$ and where $\alpha$ is a normalization factor left free in the fitting procedure. 

\begin{figure}[ht!]
 \centering
 \includegraphics[width=0.47\textwidth
]{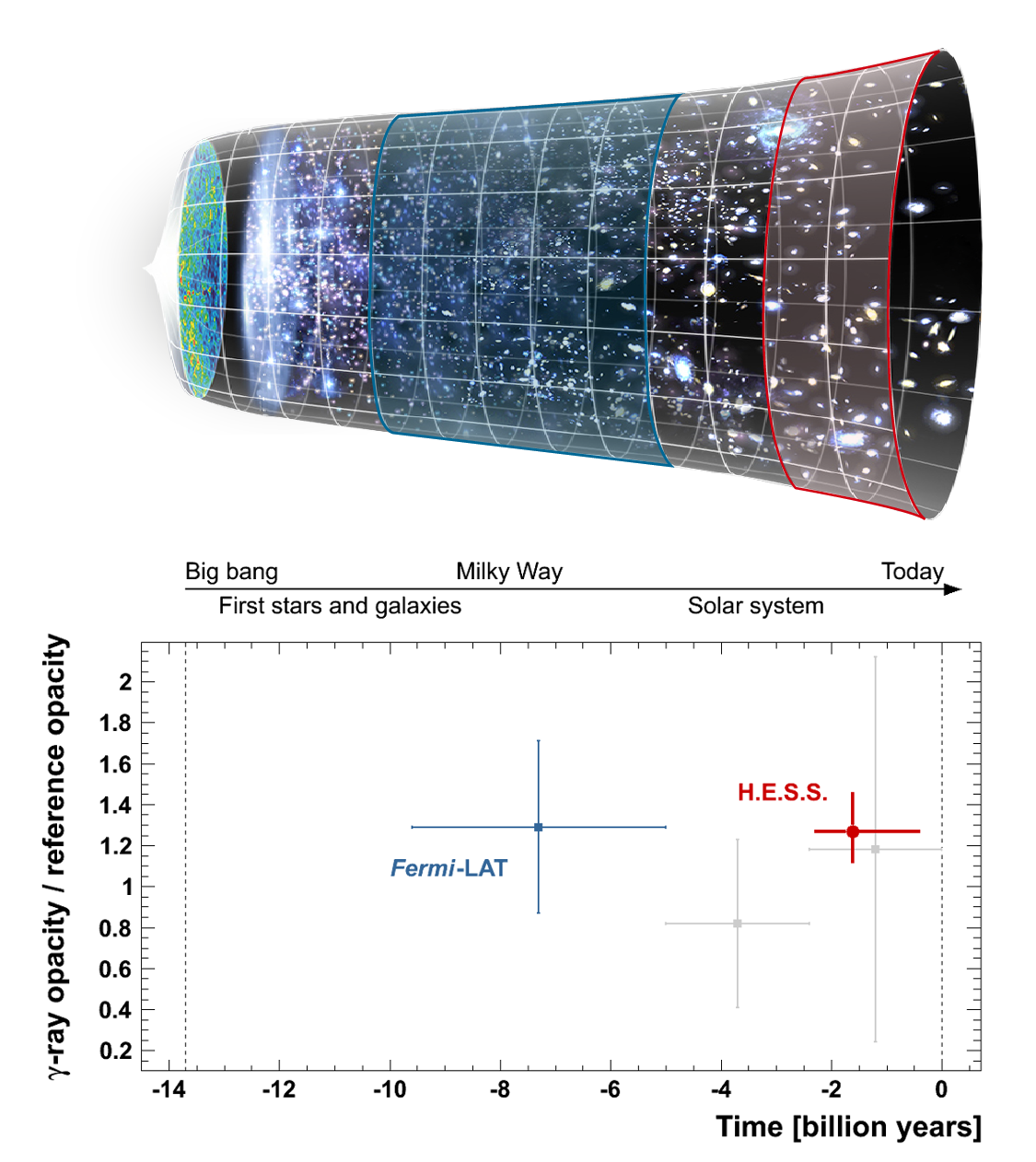}%
 \hfill \includegraphics[width=0.52\textwidth]{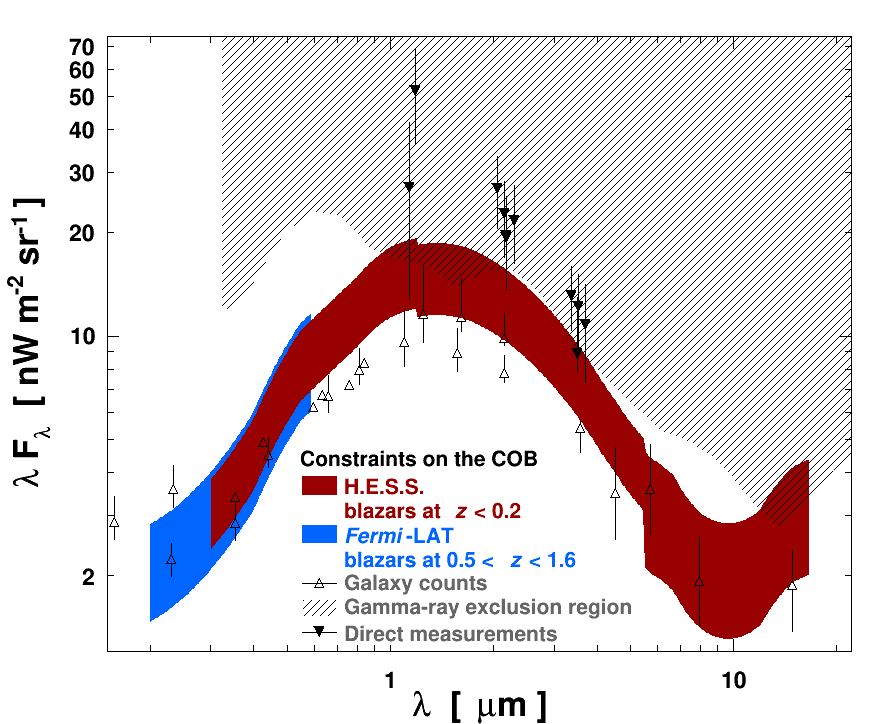}%
  \caption{{\bf Left:} EBL normalization factor as a function of the distance of the gamma-ray blazars used as beacons. The blue and gray points represent the \fermi\ measurement and constraints. The red point represents the \hess\ measurement. {\it Top graphic adapted from an original from NASA/WMAP Science team}. {\bf Right:} Spectral energy distribution of the EBL as measured by \hess\ (red) and \fermi\ (blue) from UV to near infrared wavelengths ($1\sigma$ confidence contours). {\it Adapted from} \cite{2013A&A...550A...4H}.}
  \label{Biteau:fig4}
\end{figure}

A log-likelihood profile as a function of $\alpha$ is derived for each source, the sum of these profiles resulting in the combined constraint. Using the model of \citet{Fr08} as a template for the EBL density, a detection (i.e. a normalization factor $\alpha$ differing from zero) at the $5\sigma$ level is performed with 50 sources in the redshift range $0.5<z<1.6$. Because of the limited gamma-ray energy, the EBL effect remains weak for $z<0.2$ and $0.2<z<0.5$ ($2\sigma$ measurements shown in gray in Fig.~\ref{Biteau:fig4}, left). One of the strengths of the \fermi\ study lies in the investigation of a dozen models, showing that the models of \cite{Fr08}, \cite{Dominguez} or \cite{Gilmore} fit rather well the data, despite of their completely different theoretical grounds. One of the weaknesses could lie in the extrapolation of the low-energy flux, with a fixed intrinsic emission that could result in an overestimation of the significance of the effect\footnote{The limited energy range in which the extrapolation is performed counterweights a bit this caveat.}.

The \hess\ collaboration measured the EBL absorption with observations of the seven brightest blazars ($z<0.2$) detected at VHE in the southern hemisphere \citep{2013A&A...550A...4H}. The analysis is also based on likelihood profiles as a function of a normalization factor $\alpha$. One of the weaknesses of the approach lies in the test of a single EBL model \citep{Fr08}, though retrospectively validated by the study of \fermi. Its strength lies in the treatment of the intrinsic spectrum, with the test of five different models with free parameters (even allowing for upward going intrinsic emission). The particular behavior of the EBL absorption between 1~and~$\unit[10]{TeV}$, due to the depletion of the target EBL photon field between 1~and~$\unit[10]{\mu m}$, constitutes a prominent signature that is detected by a combined likelihood analysis at the $9\sigma$ level. The normalization factor of the EBL model as measured by \hess\ and \fermi\ are compared in Fig.~\ref{Biteau:fig4}, left, the top graphic showing the epochs probed for the optical depth. The redshift range $0.2<z<0.5$ remains poorly constrained.

Scaling up the template EBL density and accounting for the energy range probed, as in \cite{2013A&A...550A...4H}, results in $1\sigma$ confidence contours shown in Fig.~\ref{Biteau:fig4}, right, where statistical and systematic uncertainties are included. The measurements lie in between lower limits from galaxy counts and upper limits from direct measurements as well as below the region excluded by the gamma-ray analyses discussed in Sec.~\ref{Sec:History}. The measurements are dominated by systematic uncertainties below $\unit[5]{\mu m}$, with an uncertainty of the order of 20-30\%. At higher wavelengths where statistical uncertainties remain dominant, there is room for improvement using larger samples of gamma rays at the highest energies.

\subsection{Constraints on the IGMF}\label{Sec:IGMFconstraints}

\begin{figure}[ht!]
 \centering
 \begin{minipage}[c]{0.48\textwidth}
	\vspace{0.42cm}
 \includegraphics[width=\textwidth]{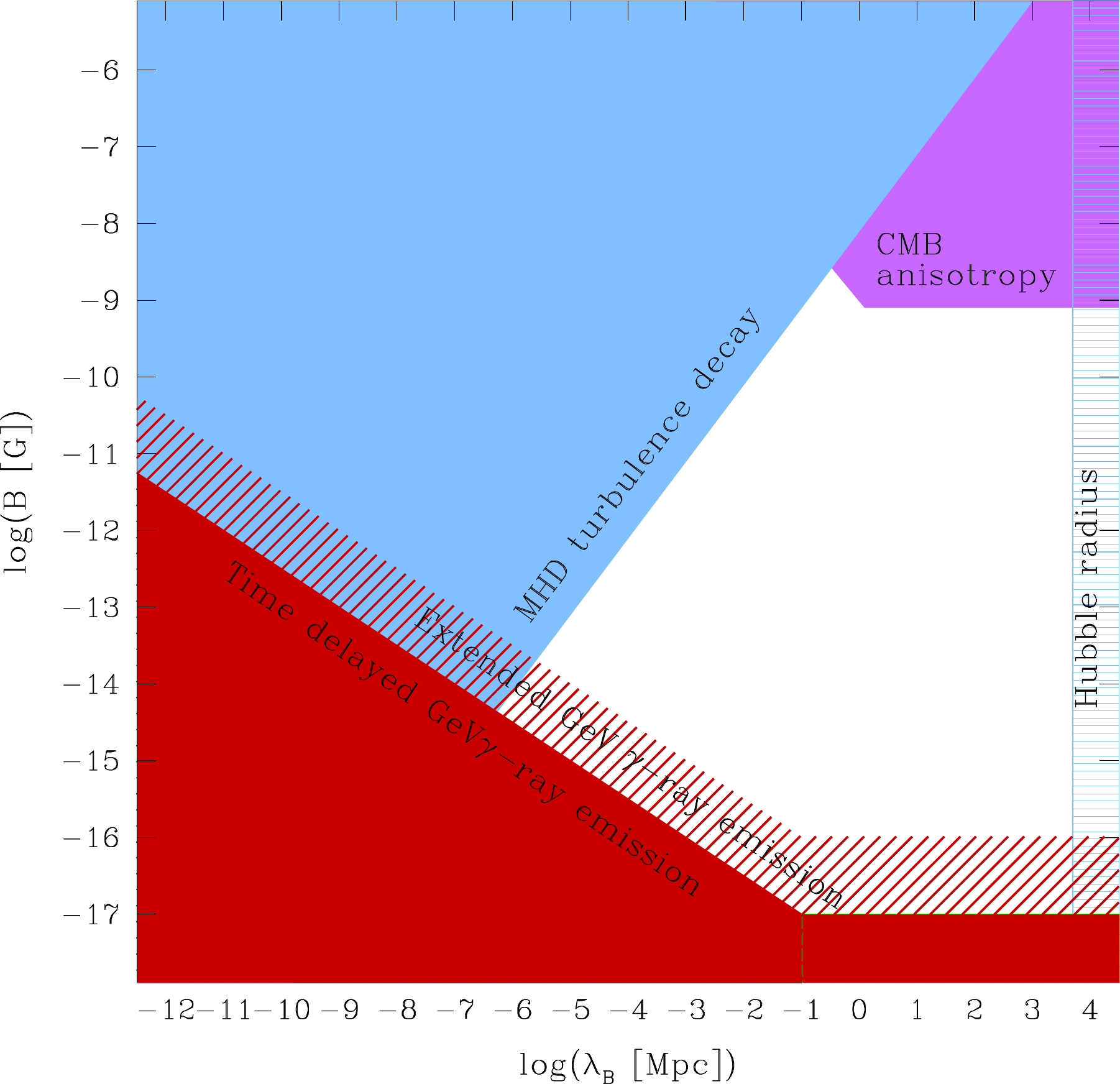}%
 \end{minipage}\hfill
 \begin{minipage}[c]{0.51\textwidth}
 \includegraphics[width=\textwidth]{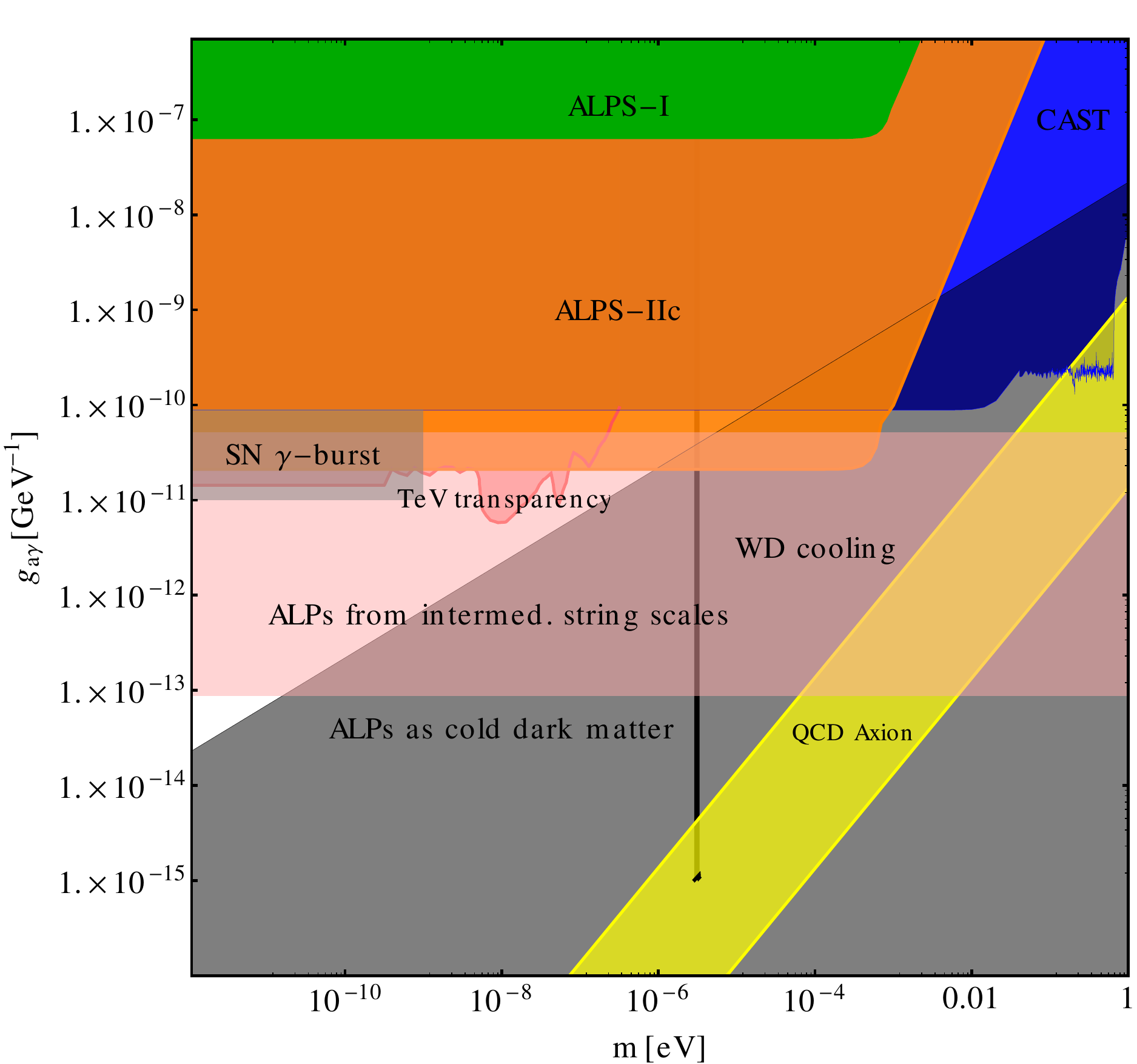}%
 \end{minipage}
  \caption{{\bf Left:} Excluded regions in parameter space intensity vs coherence length of the IGMF. {\it Adapted from } \cite{2013A&ARv..21...62D}. {\bf Right:} Constraints in the parameter space coupling of ALPs with photons vs mass. {\it Extracted from } \cite{2013JInst...8.9001B}.}
  \label{Biteau:fig6}
\end{figure}

The IGMF is usually characterized by its intensity, $B$, and its coherence length, $\lambda_B$. As discussed in \cite{2013A&ARv..21...62D}, the parameter space remains poorly constrained. A strong IGMF coherent on supra-Mpc scales would result in CMB anisotropies extending on large angular scales, resulting in an upper limit of the order $B<\unit[10^{-9}]{G}$. Large coherence lengths being untestable above the Hubble scale, the parameter space above $\lambda_B>\unit[10^{4}]{Mpc}$ is left out of the search. On theoretical grounds, MHD turbulence tends to dissipate strong magnetic fields on small scales, heating up the medium, which yields the exclusion region shown in Fig.~\ref{Biteau:fig6}, left.

As discussed in Sec.~\ref{Sec:IGMFth}, joint observations of gamma-ray sources at HE and VHE can probe the intensity of the IGMF. For low enough $B$, the EBL-absorption induced pairs should barely be deflected, resulting in a reprocessing of the signal to HE. The non-observation of this signal, mostly in the spectral energy distribution of the HSP blazar 1ES~0229+200, results in a lower limit of the order of $B>\unit[10^{-16}]{G}$ \citep[][shown as a dashed red region in Fig.~\ref{Biteau:fig6}, left]{2010Sci...328...73N}, under the assumption of a fixed EBL model and of a longterm steady-state emission of the source. Releasing the latter assumption, \cite{2011ApJ...733L..21D} and \cite{2011A&A...529A.144T} derived weaker limits of the order $\unit[10^{-18}]{G}$ and $\unit[10^{-17}]{G}$, the latter being shown as a filled region in Fig.~\ref{Biteau:fig6}, left. Investigating the underlying hypotheses on the source intrinsic emission, \cite{2012arXiv1210.2802A} concluded than present data on 1ES~0229+200 rule the null IGMF hypothesis out at the $3\sigma$ confidence level. The strong dependence of these constraints on the EBL level is discussed in \cite{2012ApJ...747L..14V}, with an increase in the lower limit of almost a factor of 10 for an increase in the EBL level by 30\%, which is representative of the current uncertainties derived by \cite{2013A&A...550A...4H} and \cite{2012Sci...338.1190A}. 

\subsection{Hints of / limits on physics beyond the standard models}\label{Sec:BSMhints}

Some puzzles remain to be solved at VHE. For example, the spectrum of the blazar PKS~1424+240 measured by the VERITAS collaboration does not seem as absorbed as it should be, given the redshift of the source $z\geq0.6035$ \citep[inferred from Ly$\beta$ and Ly$\gamma$ absorption in][]{2013ApJ...768L..31F}. Other hints for deviations from nominal EBL opacities have recently been studied in \cite{2012JCAP...02..033H} using VHE spectral points. Archival points are fitted with the model described in Eq.~\ref{Eq:SpecModel}, assuming a power-law or a log-parabolic intrinsic spectrum and using a template EBL model with a normalization factor of one. An estimator of the residual to the fit is defined for each point as:

\begin{equation}
R(E_i) = {{F_i - \phi(E_i)} \over {F_i + \phi(E_i)}}
\label{Eq:estimator}
\end{equation}

\noindent where $F_i$ is the flux measured at the energy $E_i$. The authors discard an estimator taking into account the measured uncertainties on the flux, such as $({F_i - \phi(E_i))/{\sigma_i}}$, because of deviations from the expected normal distribution of unity width.
At high optical depth, or equivalently at high energies for a given redshift, a $4\sigma$ deviation from null residuals is found, indicating an opacity smaller than expected. Systematic uncertainties in the energy scale decrease the effect to the $3.5\sigma$ level, and discarding the last points from each spectrum in the study reduces the deviation to the $2.5\sigma$ level. In \cite{2013arXiv1309.3846H}, a similar approach is followed using three \fermi\ events from a gamma-ray burst and two blazars, yielding a $3.5\sigma$ effect (preliminary value).

The VHE hint has been used by the same group in \cite{2013PhRvD..87c5027M} to constrain the parameter space of ALPs, i.e their coupling to photons as a function of the ALP mass, and claim the first lower-limit on the coupling (complex-shaped dark pink region in Fig.~\ref{Biteau:fig6}, right). The region of interest, denoted as ``TeV transparency'', lies below the regions excluded by the CAST and ALPS-I experiments and is partially excluded by the signal from SN~1987A. It intersects the light pink band possibly explaining anomalous cooling rates from white dwarves, but remains above theoretical expectations from string theory, cold dark matter scenarios, or standard axions invoked as a solution to the absence of CP-violation in QCD, which are in turn probed by the ADMX experiment between 2 and $\unit[3]{\mu eV}$. Not included in this graphic is the recent result from \cite{2013arXiv1307.6068B} that excludes a large fraction of the ``TeV transparency'' region using VHE data from the HSP blazar PKS~2155-304.

As exciting as these hints could be, some systematic uncertainties could affect the estimated significance level of the effect. The EBL model is subject to large uncertainties in the energy range concerned (typically above $\unit[5]{\mu m}$). For the \fermi\ event analysis, the method does not seem to account for the energy resolution of the instruments. For the analysis of VHE spectra, the estimator described in Eq.~\ref{Eq:estimator} does not account for the uncertainties on the flux measurements, which drastically increase with energy. Fitting models to published spectral points, though common practice, does not properly account for measurement uncertainties and correlations between points\footnote{VHE points are generally highly correlated due to the steepness of the spectra and the limited energy resolution that imply leakage of events from low to higher energies. Spectral points derived with a maximum likelihood analysis should be considered as residuals to the fit. Therefore, fitting a model to these residuals may not be the most statistically sound approach.}. The last caveat, briefly addressed in the study of systematic uncertainties by \cite{2012JCAP...02..033H}, concerns the highest energy points in each spectrum. The last point above some minimum significance is indeed usually selected when publishing a VHE spectrum, thus biasing the selection toward upward fluctuations with respect to the actual flux. Suppressing these points largely reduces the significance of the effect (down to $2.5\sigma$), potentially suggesting an accumulation of upward going fluctuations. It could also be argued that these high-energy points are precisely the ones from which the ALP signal is expected. Along with laboratory experiments such as ALPS-II, large improvements in the energy coverage, discussed in Sec.~\ref{Sec:Future}, are needed from future generation VHE instruments to decide this issue.

\subsection{EBL absorption as a ruler?}

From Eq.~\ref{Eq:EBLOptDep}, the EBL optical depth first appears as a probe for deviations from the pair-creation cross section (third integral in the equation), as discussed in Sec.~\ref{Sec:BSMth} with Lorentz invariance violation, then as a proxy for the EBL density (second integral), as discussed in Sec.~\ref{Sec:EBLmeasurements} with the measurements of \hess\ and \fermi. Alternatively, it also could in principle be used to constrain distances (first integral). 

Upper limits on the redshifts of HSP blazars, such as PG~1553+113 \citep[see, e.g.,][]{2006A&A...448L..19A}, have been set on a regular basis, going hand in hand with growingly stringent limits on the EBL. Attempts to standardize the multiwavelength emission of blazars at HE and VHE have been led by \cite{2010MNRAS.405L..76P}, correlating the HE-VHE spectral differences, which trace EBL absorption, with distance. The authors infer a redshift of $z = 0.24 \pm 0.05$ for PKS~1424+240 based on the \veritas\ and \fermi\ spectral measurements, emphasizing the puzzle around this object (cf. Sec.~\ref{Sec:BSMhints}). 

Constraints on the Hubble parameter $H_0$ have also been derived using the absorption of gamma-rays from blazars \citep{2008MNRAS.389..919B,2013ApJ...771L..34D}. Note however that even for an EBL optical depth virtually perfectly measured, the remaining uncertainties on the EBL density (20-30\%) would be directly propagated into the inferred distances and Hubble parameter. 

\section{Perspectives in gamma-ray cosmology}\label{Sec:Future} 

\subsection{Using the multi-wavelength synergy}

The constraints derived from gamma-ray cosmology are based on a joint modeling of the source emission, EBL absorption and second order processes, such as the effect of the IGMF or of physics beyond the standard models. Using the spectra derived during multi-wavelength campaigns in the X-ray, HE and VHE gamma-ray bands has proved particularly constraining for the modeling of blazars' emission. Using this multi-wavelength synergy, in particular the evolution of HE-VHE spectra with redshift, \cite{2013A&A...554A..75S}  have recently confirmed the \hess\ and \fermi\ measurements. 

Using in addition X-ray observations, \cite{2010ApJ...715L..16M} and \cite{2013ApJ...770...77D} modeled the broad band spectra of blazars within synchrotron self-Compton scenarios and jointly constrained the EBL optical depth, though the same caveat as in Sec.~\ref{Sec:BSMhints} (uncertainties and correlations between the points) could be raised. The multi-wavelength approach is crucial for probes of the IGMF based on pair-echo, with the reprocessing of the VHE signal to HE. The IGMF intensity could nonetheless potentially be constrained up to $\unit[10^{-12}]{G}$ solely with CTA data, using pair-haloes / beam-broadened-cascades approaches \citep{2013A&ARv..21...62D,2013IAUS..294..459S}.

\subsection{Scope of current generation instruments}

The lowering of the energy threshold of ground-based gamma-ray instruments below $\unit[100]{GeV}$ is a crucial step in widening the sample of probes of the Universe opacity to gamma rays. To increase the sensitivity to Cherenkov light from low-energy gamma-ray induced atmospheric showers, the VERITAS collaboration opted for high quantum efficiency photomultipliers, while the MAGIC and H.E.S.S. collaborations opted for large size telescopes (28~m diameter for the recently built CT5 telescope of the H.E.S.S.~II array). A low energy threshold implies a larger gamma-ray horizon, as shown e.g. in Fig.~\ref{Biteau:fig2b}. VHE instruments could then probe the EBL optical depth using blazars at $0.2<z<0.5$, which remains uncharted territory (cf. Fig.~\ref{Biteau:fig4}, left). Measuring gamma-ray opacity in various redshift bands would result in a tomography of the EBL density, enabling the comparison of the EBL evolution with e.g. the star formation history \citep{2012AIPC.1505..610R}.

\subsection{Gamma-ray cosmology with CTA}

Based on the \fermi\ population of extragalactic sources, the population detected by ground-based instruments should become less and less dominated by HSP objects, allowing tests of unification schemes of blazars. These investigations, which also constitute a fundamental science case for CTA \citep[see][and references therein]{2013APh....43..103R,2013APh....43..215S}, will deepen our knowledge on potential beacons and thus directly benefit gamma-ray cosmology. The knowledge of the Universe opacity at VHE will also directly impact the physics of gamma-ray bursts \citep[see][in the context of CTA]{2013APh....43..252I,2013APh....43..134M}, which should be detected at a rate of a few per year by CTA \citep{2013ExA....35..413G}.

CTA, with its large-size telescopes enabling an energy threshold of $\unit[30]{GeV}$ or less, its medium-size telescopes boosting the sensitivity by a factor of 10 in the core energy range, and small size telescopes paving a few kilometer-square area that will probe energies above $\unit[10]{TeV}$, will take gamma-ray cosmology to its full maturity \citep[see][for discussions on EBL and Lorentz invariance in the CTA era]{2013APh....43..241M,2013APh....43...50E}. CTA will in particular probe the EBL density above $\unit[5]{\mu m}$, where diffuse emission from polycyclic aromatic hydrocarbons could be detected and where physics beyond the standard models could be unveiled, yielding observables for quantum gravity theories or clues on the elusive question of the nature of dark matter.

\begin{acknowledgements}
JB would like to thank D. Williams for his useful suggestions, which improved this text.
\end{acknowledgements}

\bibliographystyle{aa}  
\bibliography{Biteau.bbl} 

\begin{thebibliography}{68}
\expandafter\ifx\csname natexlab\endcsname\relax\def\natexlab#1{#1}\fi

\bibitem[{{Abdo} {et~al.}(2010){Abdo}, {Ackermann}, {Ajello}, {Allafort},
  {Atwood}, {Baldini}, {Ballet}, {Barbiellini}, {Baring}, {Bastieri},
  {Baughman}, {Bechtol}, {Bellazzini}, {Berenji}, {Bhat}, {Blandford}, {Bloom},
  {Bonamente}, {Borgland}, {Bouvier}, {Brandt}, {Bregeon}, {Brez}, {Briggs},
  {Brigida}, {Bruel}, {Buehler}, {Burnett}, {Buson}, {Caliandro}, {Cameron},
  {Caraveo}, {Carrigan}, {Casandjian}, {Cavazzuti}, {Cecchi}, {{\c C}elik},
  {Charles}, {Chekhtman}, {Chen}, {Cheung}, {Chiang}, {Ciprini}, {Claus},
  {Cohen-Tanugi}, {Connaughton}, {Conrad}, {Costamante}, {Dermer}, {de
  Angelis}, {de Palma}, {Digel}, {Dingus}, {Silva}, {Drell}, {Dubois},
  {Favuzzi}, {Fegan}, {Finke}, {Fortin}, {Fukazawa}, {Funk}, {Fusco},
  {Gargano}, {Gasparrini}, {Gehrels}, {Germani}, {Giglietto}, {Gilmore},
  {Giommi}, {Giordano}, {Giroletti}, {Glanzman}, {Godfrey}, {Granot},
  {Greiner}, {Grenier}, {Grove}, {Guiriec}, {Gustafsson}, {Hadasch},
  {Hayashida}, {Hays}, {Horan}, {Hughes}, {J{\'o}hannesson}, {Johnson},
  {Johnson}, {Johnson}, {Kamae}, {Katagiri}, {Kataoka}, {Kn{\"o}dlseder},
  {Kocevski}, {Kuss}, {Lande}, {Latronico}, {Lee}, {Llena Garde}, {Longo},
  {Loparco}, {Lott}, {Lovellette}, {Lubrano}, {Makeev}, {Mazziotta},
  {McConville}, {McEnery}, {McGlynn}, {Mehault}, {M{\'e}sz{\'a}ros},
  {Michelson}, {Mizuno}, {Moiseev}, {Monte}, {Monzani}, {Moretti}, {Morselli},
  {Moskalenko}, {Murgia}, {Nakamori}, {Naumann-Godo}, {Nolan}, {Norris},
  {Nuss}, {Ohno}, {Ohsugi}, {Okumura}, {Omodei}, {Orlando}, {Ormes}, {Ozaki},
  {Paneque}, {Panetta}, {Parent}, {Pelassa}, {Pepe}, {Pesce-Rollins}, {Piron},
  {Porter}, {Primack}, {Rain{\`o}}, {Rando}, {Razzano}, {Razzaque}, {Reimer},
  {Reimer}, {Reyes}, {Ripken}, {Ritz}, {Romani}, {Roth}, {Sadrozinski},
  {Sanchez}, {Sander}, {Scargle}, {Schalk}, {Sgr{\`o}}, {Shaw}, {Siskind},
  {Smith}, {Spandre}, {Spinelli}, {Stecker}, {Strickman}, {Suson}, {Tajima},
  {Takahashi}, {Takahashi}, {Tanaka}, {Thayer}, {Thayer}, {Thompson},
  {Tibaldo}, {Torres}, {Tosti}, {Tramacere}, {Uchiyama}, {Usher},
  {Vandenbroucke}, {Vasileiou}, {Vilchez}, {Vitale}, {von Kienlin}, {Waite},
  {Wang}, {Wilson-Hodge}, {Winer}, {Wood}, {Yamazaki}, {Yang}, {Ylinen}, \&
  {Ziegler}}]{UpperLimitFermi}
{Abdo}, A.~A., {Ackermann}, M., {Ajello}, M., {et~al.} 2010, \apj, 723, 1082

\bibitem[{{Ackermann} {et~al.}(2012){Ackermann}, {Ajello}, {Allafort},
  {Schady}, {Baldini}, {Ballet}, {Barbiellini}, {Bastieri}, {Bellazzini},
  {Blandford}, {Bloom}, {Borgland}, {Bottacini}, {Bouvier}, {Bregeon},
  {Brigida}, {Bruel}, {Buehler}, {Buson}, {Caliandro}, {Cameron}, {Caraveo},
  {Cavazzuti}, {Cecchi}, {Charles}, {Chaves}, {Chekhtman}, {Cheung}, {Chiang},
  {Chiaro}, {Ciprini}, {Claus}, {Cohen-Tanugi}, {Conrad}, {Cutini},
  {D'Ammando}, {de Palma}, {Dermer}, {Digel}, {do Couto e Silva},
  {Dom{\'{\i}}nguez}, {Drell}, {Drlica-Wagner}, {Favuzzi}, {Fegan}, {Focke},
  {Franckowiak}, {Fukazawa}, {Funk}, {Fusco}, {Gargano}, {Gasparrini},
  {Gehrels}, {Germani}, {Giglietto}, {Giordano}, {Giroletti}, {Glanzman},
  {Godfrey}, {Grenier}, {Grove}, {Guiriec}, {Gustafsson}, {Hadasch},
  {Hayashida}, {Hays}, {Jackson}, {Jogler}, {Kataoka}, {Kn{\"o}dlseder},
  {Kuss}, {Lande}, {Larsson}, {Latronico}, {Longo}, {Loparco}, {Lovellette},
  {Lubrano}, {Mazziotta}, {McEnery}, {Mehault}, {Michelson}, {Mizuno}, {Monte},
  {Monzani}, {Morselli}, {Moskalenko}, {Murgia}, {Tramacere}, {Nuss},
  {Greiner}, {Ohno}, {Ohsugi}, {Omodei}, {Orienti}, {Orlando}, {Ormes},
  {Paneque}, {Perkins}, {Pesce-Rollins}, {Piron}, {Pivato}, {Porter},
  {Rain{\`o}}, {Rando}, {Razzano}, {Razzaque}, {Reimer}, {Reimer}, {Reyes},
  {Ritz}, {Rau}, {Romoli}, {Roth}, {S{\'a}nchez-Conde}, {Sanchez}, {Scargle},
  {Sgr{\`o}}, {Siskind}, {Spandre}, {Spinelli}, {Stawarz}, {Suson},
  {Takahashi}, {Tanaka}, {Thayer}, {Thompson}, {Tibaldo}, {Tinivella},
  {Torres}, {Tosti}, {Troja}, {Usher}, {Vandenbroucke}, {Vasileiou},
  {Vianello}, {Vitale}, {Waite}, {Winer}, {Wood}, \&
  {Wood}}]{2012Sci...338.1190A}
{Ackermann}, M., {Ajello}, M., {Allafort}, A., {et~al.} 2012, Science, 338,
  1190

\bibitem[{{Aharonian} {et~al.}(2006{\natexlab{a}}){Aharonian}, {Akhperjanian},
  {Bazer-Bachi}, {Beilicke}, {Benbow}, {Berge}, {Bernl{\"o}hr}, {Boisson},
  {Bolz}, {Borrel}, {Braun}, {Breitling}, {Brown}, {B{\"u}hler}, {Carrigan},
  {Chadwick}, {Chounet}, {Cornils}, {Costamante}, {Degrange}, {Dickinson},
  {Djannati-Ata{\"i}}, {O'C.~Drury}, {Dubus}, {Egberts}, {Emmanoulopoulos},
  {Espigat}, {Feinstein}, {Fontaine}, {Funk}, {Gallant}, {Giebels},
  {Glicenstein}, {Goret}, {Hadjichristidis}, {Hauser}, {Hauser}, {Heinzelmann},
  {Henri}, {Hermann}, {Hinton}, {Hofmann}, {Holleran}, {Horns}, {Jacholkowska},
  {de Jager}, {Kh{\'e}lifi}, {Komin}, {Konopelko}, {Latham}, {Le Gallou},
  {Lemi{\`e}re}, {Lemoine-Goumard}, {Lohse}, {Martin}, {Martineau-Huynh},
  {Marcowith}, {Masterson}, {McComb}, {de Naurois}, {Nedbal}, {Nolan},
  {Noutsos}, {Orford}, {Osborne}, {Ouchrif}, {Panter}, {Pelletier}, {Pita},
  {P{\"u}hlhofer}, {Punch}, {Raubenheimer}, {Raue}, {Rayner}, {Reimer},
  {Reimer}, {Ripken}, {Rob}, {Rolland}, {Rowell}, {Sahakian}, {Saug{\'e}},
  {Schlenker}, {Schlickeiser}, {Schuster}, {Schwanke}, {Siewert}, {Sol},
  {Spangler}, {Steenkamp}, {Stegmann}, {Superina}, {Tavernet}, {Terrier},
  {Th{\'e}oret}, {Tluczykont}, {van Eldik}, {Vasileiadis}, {Venter}, {Vincent},
  {V{\"o}lk}, {Wagner}, \& {Ward}}]{2006A&A...448L..19A}
{Aharonian}, F., {Akhperjanian}, A.~G., {Bazer-Bachi}, A.~R., {et~al.}
  2006{\natexlab{a}}, \aap, 448, L19

\bibitem[{{Aharonian} {et~al.}(2006{\natexlab{b}}){Aharonian}, {Akhperjanian},
  {Bazer-Bachi}, {Beilicke}, {Benbow}, {Berge}, {Bernl{\"o}hr}, {Boisson},
  {Bolz}, {Borrel}, {Braun}, {Breitling}, {Brown}, {Chadwick}, {Chounet},
  {Cornils}, {Costamante}, {Degrange}, {Dickinson}, {Djannati-Ata{\"i}},
  {Drury}, {Dubus}, {Emmanoulopoulos}, {Espigat}, {Feinstein}, {Fontaine},
  {Fuchs}, {Funk}, {Gallant}, {Giebels}, {Gillessen}, {Glicenstein}, {Goret},
  {Hadjichristidis}, {Hauser}, {Hauser}, {Heinzelmann}, {Henri}, {Hermann},
  {Hinton}, {Hofmann}, {Holleran}, {Horns}, {Jacholkowska}, {de Jager},
  {Kh{\'e}lifi}, {Klages}, {Komin}, {Konopelko}, {Latham}, {Le Gallou},
  {Lemi{\`e}re}, {Lemoine-Goumard}, {Leroy}, {Lohse}, {Martin},
  {Martineau-Huynh}, {Marcowith}, {Masterson}, {McComb}, {de Naurois}, {Nolan},
  {Noutsos}, {Orford}, {Osborne}, {Ouchrif}, {Panter}, {Pelletier}, {Pita},
  {P{\"u}hlhofer}, {Punch}, {Raubenheimer}, {Raue}, {Raux}, {Rayner}, {Reimer},
  {Reimer}, {Ripken}, {Rob}, {Rolland}, {Rowell}, {Sahakian}, {Saug{\'e}},
  {Schlenker}, {Schlickeiser}, {Schuster}, {Schwanke}, {Siewert}, {Sol},
  {Spangler}, {Steenkamp}, {Stegmann}, {Tavernet}, {Terrier}, {Th{\'e}oret},
  {Tluczykont}, {van Eldik}, {Vasileiadis}, {Venter}, {Vincent}, {V{\"o}lk}, \&
  {Wagner}}]{EBLAHA}
{Aharonian}, F., {Akhperjanian}, A.~G., {Bazer-Bachi}, A.~R., {et~al.}
  2006{\natexlab{b}}, \nat, 440, 1018

\bibitem[{{Aharonian} {et~al.}(1999){Aharonian}, {Akhperjanian}, {Barrio},
  {Bernl{\"o}hr}, {Bojahr}, {Calle}, {Contreras}, {Cortina}, {Daum}, {Deckers},
  {Denninghoff}, {Fonseca}, {Gonzalez}, {Heinzelmann}, {Hemberger}, {Hermann},
  {He{\ss}}, {Heusler}, {Hofmann}, {Hohl}, {Horns}, {Ibarra}, {Kankanyan},
  {Kettler}, {K{\"o}hler}, {Konopelko}, {Kornmeyer}, {Kestel}, {Kranich},
  {Krawczynski}, {Lampeitl}, {Lindner}, {Lorenz}, {Magnussen}, {Meyer},
  {Mirzoyan}, {Moralejo}, {Padilla}, {Panter}, {Petry}, {Plaga},
  {Plyasheshnikov}, {Prahl}, {P{\"u}hlhofer}, {Rauterberg}, {Renault}, {Rhode},
  {R{\"o}hring}, {Sahakian}, {Samorski}, {Schmele}, {Schr{\"o}der}, {Stamm},
  {V{\"o}lk}, {Wiebel-Sooth}, {Wiedner}, {Willmer}, \&
  {Wittek}}]{1999A&A...349...11A}
{Aharonian}, F.~A., {Akhperjanian}, A.~G., {Barrio}, J.~A., {et~al.} 1999,
  \aap, 349, 11

\bibitem[{{Antonucci}(1993)}]{1993ARA&A..31..473A}
{Antonucci}, R. 1993, \araa, 31, 473

\bibitem[{{Arlen} {et~al.}(2012){Arlen}, {Vassiliev}, {Weisgarber}, {Wakely},
  \& {Yusef Shafi}}]{2012arXiv1210.2802A}
{Arlen}, T.~C., {Vassiliev}, V.~V., {Weisgarber}, T., {Wakely}, S.~P., \&
  {Yusef Shafi}, S. 2012, arXiv:1210.2802

\bibitem[{{B{\"a}hre} {et~al.}(2013){B{\"a}hre}, {D{\"o}brich},
  {Dreyling-Eschweiler}, {Ghazaryan}, {Hodajerdi}, {Horns}, {Januschek},
  {Knabbe}, {Lindner}, {Notz}, {Ringwald}, {von Seggern}, {Stromhagen},
  {Trines}, \& {Willke}}]{2013JInst...8.9001B}
{B{\"a}hre}, R., {D{\"o}brich}, B., {Dreyling-Eschweiler}, J., {et~al.} 2013,
  Journal of Instrumentation, 8, 9001

\bibitem[{{Band} \& {Grindlay}(1985)}]{1985ApJ...298..128B}
{Band}, D.~L. \& {Grindlay}, J.~E. 1985, \apj, 298, 128

\bibitem[{{Barrau} {et~al.}(2008){Barrau}, {Gorecki}, \&
  {Grain}}]{2008MNRAS.389..919B}
{Barrau}, A., {Gorecki}, A., \& {Grain}, J. 2008, \mnras, 389, 919

\bibitem[{{Biller} {et~al.}(1998){Biller}, {Buckley}, {Burdett}, {Bussons
  Gordo}, {Carter-Lewis}, {Fegan}, {Finley}, {Gaidos}, {Hillas}, {Krennrich},
  {Lamb}, {Lessard}, {McEnery}, {Mohanty}, {Quinn}, {Rodgers}, {Rose},
  {Samuelson}, {Sembroski}, {Skelton}, {Weekes}, \&
  {Zweerink}}]{1998PhRvL..80.2992B}
{Biller}, S.~D., {Buckley}, J., {Burdett}, A., {et~al.} 1998, Physical Review
  Letters, 80, 2992

\bibitem[{{Broderick} {et~al.}(2012){Broderick}, {Chang}, \&
  {Pfrommer}}]{2012ApJ...752...22B}
{Broderick}, A.~E., {Chang}, P., \& {Pfrommer}, C. 2012, \apj, 752, 22

\bibitem[{{Brun} {et~al.}(2013){Brun}, {Wouters}, \& {for the
  H.~E.~S.~S.~collaboration}}]{2013arXiv1307.6068B}
{Brun}, P., {Wouters}, D., \& {for the H.~E.~S.~S.~collaboration}. 2013,
  arXiv:1307.6068

\bibitem[{{Dermer} {et~al.}(2011){Dermer}, {Cavadini}, {Razzaque}, {Finke},
  {Chiang}, \& {Lott}}]{2011ApJ...733L..21D}
{Dermer}, C.~D., {Cavadini}, M., {Razzaque}, S., {et~al.} 2011, \apjl, 733, L21

\bibitem[{{Dole} {et~al.}(2006){Dole}, {Lagache}, {Puget}, {Caputi},
  {Fern{\'a}ndez-Conde}, {Le Floc'h}, {Papovich}, {P{\'e}rez-Gonz{\'a}lez},
  {Rieke}, \& {Blaylock}}]{Dole}
{Dole}, H., {Lagache}, G., {Puget}, J.-L., {et~al.} 2006, \aap, 451, 417

\bibitem[{{Dom{\'{\i}}nguez} {et~al.}(2013){Dom{\'{\i}}nguez}, {Finke},
  {Prada}, {Primack}, {Kitaura}, {Siana}, \& {Paneque}}]{2013ApJ...770...77D}
{Dom{\'{\i}}nguez}, A., {Finke}, J.~D., {Prada}, F., {et~al.} 2013, \apj, 770,
  77

\bibitem[{{Dom{\'{\i}}nguez} \& {Prada}(2013)}]{2013ApJ...771L..34D}
{Dom{\'{\i}}nguez}, A. \& {Prada}, F. 2013, \apjl, 771, L34

\bibitem[{{Dom{\'{\i}}nguez} {et~al.}(2011){Dom{\'{\i}}nguez}, {Primack},
  {Rosario}, {Prada}, {Gilmore}, {Faber}, {Koo}, {Somerville},
  {P{\'e}rez-Torres}, {P{\'e}rez-Gonz{\'a}lez}, {Huang}, {Davis},
  {Guhathakurta}, {Barmby}, {Conselice}, {Lozano}, {Newman}, \&
  {Cooper}}]{Dominguez}
{Dom{\'{\i}}nguez}, A., {Primack}, J.~R., {Rosario}, D.~J., {et~al.} 2011,
  \mnras, 410, 2556

\bibitem[{{Durrer} \& {Neronov}(2013)}]{2013A&ARv..21...62D}
{Durrer}, R. \& {Neronov}, A. 2013, \aapr, 21, 62

\bibitem[{{Dwek} \& {Krennrich}(2005)}]{Dwek}
{Dwek}, E. \& {Krennrich}, F. 2005, \apj, 618, 657

\bibitem[{{Dwek} \& {Krennrich}(2013)}]{2013APh....43..112D}
{Dwek}, E. \& {Krennrich}, F. 2013, Astroparticle Physics, 43, 112

\bibitem[{{Dwek} \& {Slavin}(1994)}]{1994ApJ...436..696D}
{Dwek}, E. \& {Slavin}, J. 1994, \apj, 436, 696

\bibitem[{{Ellis} \& {Mavromatos}(2013)}]{2013APh....43...50E}
{Ellis}, J. \& {Mavromatos}, N.~E. 2013, Astroparticle Physics, 43, 50

\bibitem[{{Fazio} {et~al.}(2004){Fazio}, {Ashby}, {Barmby}, {Hora}, {Huang},
  {Pahre}, {Wang}, {Willner}, {Arendt}, {Moseley}, {Brodwin}, {Eisenhardt},
  {Stern}, {Tollestrup}, \& {Wright}}]{2004ApJS..154...39F}
{Fazio}, G.~G., {Ashby}, M.~L.~N., {Barmby}, P., {et~al.} 2004, \apjs, 154, 39

\bibitem[{{Franceschini} {et~al.}(2008){Franceschini}, {Rodighiero}, \&
  {Vaccari}}]{Fr08}
{Franceschini}, A., {Rodighiero}, G., \& {Vaccari}, M. 2008, \aap, 487, 837

\bibitem[{{Funk} {et~al.}(1998){Funk}, {Magnussen}, {Meyer}, {Rhode},
  {Westerhoff}, \& {Wiebel-Sooth}}]{1998APh.....9...97F}
{Funk}, B., {Magnussen}, N., {Meyer}, H., {et~al.} 1998, Astroparticle Physics,
  9, 97

\bibitem[{{Furniss} {et~al.}(2013){Furniss}, {Williams}, {Danforth},
  {Fumagalli}, {Prochaska}, {Primack}, {Urry}, {Stocke}, {Filippenko}, \&
  {Neely}}]{2013ApJ...768L..31F}
{Furniss}, A., {Williams}, D.~A., {Danforth}, C., {et~al.} 2013, \apjl, 768,
  L31

\bibitem[{{Georganopoulos} {et~al.}(2010){Georganopoulos}, {Finke}, \&
  {Reyes}}]{Georganopoulos}
{Georganopoulos}, M., {Finke}, J.~D., \& {Reyes}, L.~C. 2010, \apjl, 714, L157

\bibitem[{{Ghisellini}(2013)}]{2013arXiv1309.4772G}
{Ghisellini}, G. 2013, arXiv:1309.4772

\bibitem[{{Gilmore} {et~al.}(2013){Gilmore}, {Bouvier}, {Connaughton},
  {Goldstein}, {Otte}, {Primack}, \& {Williams}}]{2013ExA....35..413G}
{Gilmore}, R.~C., {Bouvier}, A., {Connaughton}, V., {et~al.} 2013, Experimental
  Astronomy, 35, 413

\bibitem[{{Gilmore} {et~al.}(2012){Gilmore}, {Somerville}, {Primack}, \&
  {Dom{\'{\i}}nguez}}]{Gilmore}
{Gilmore}, R.~C., {Somerville}, R.~S., {Primack}, J.~R., \& {Dom{\'{\i}}nguez},
  A. 2012, \mnras, 422, 3189

\bibitem[{{Gould} \& {Schr{\'e}der}(1967)}]{1967PhRv..155.1408G}
{Gould}, R.~J. \& {Schr{\'e}der}, G.~P. 1967, Physical Review, 155, 1408

\bibitem[{{Guy} {et~al.}(2000){Guy}, {Renault}, {Aharonian}, {Rivoal}, \&
  {Tavernet}}]{2000A&A...359..419G}
{Guy}, J., {Renault}, C., {Aharonian}, F.~A., {Rivoal}, M., \& {Tavernet},
  J.-P. 2000, \aap, 359, 419

\bibitem[{{Hauser} \& {Dwek}(2001)}]{HauserDwek}
{Hauser}, M.~G. \& {Dwek}, E. 2001, \araa, 39, 249

\bibitem[{{H.E.S.S.~Collaboration} {et~al.}(2013){H.E.S.S.~Collaboration},
  {Abramowski}, {Acero}, {Aharonian}, {Akhperjanian}, {Anton}, {Balenderan},
  {Balzer}, {Barnacka}, {Becherini}, {Becker Tjus}, {Bernl{\"o}hr}, {Birsin},
  {Biteau}, {Bochow}, {Boisson}, {Bolmont}, {Bordas}, {Brucker}, {Brun},
  {Brun}, {Bulik}, {Carrigan}, {Casanova}, {Cerruti}, {Chadwick},
  {Charbonnier}, {Chaves}, {Cheesebrough}, {Cologna}, {Conrad}, {Couturier},
  {Dalton}, {Daniel}, {Davids}, {Degrange}, {Deil}, {deWilt}, {Dickinson},
  {Djannati-Ata{\"i}}, {Domainko}, {O'C.~Drury}, {Dubus}, {Dutson}, {Dyks},
  {Dyrda}, {Egberts}, {Eger}, {Espigat}, {Fallon}, {Farnier}, {Fegan},
  {Feinstein}, {Fernandes}, {Fernandez}, {Fiasson}, {Fontaine}, {F{\"o}rster},
  {F{\"u}{\ss}ling}, {Gajdus}, {Gallant}, {Garrigoux}, {Gast}, {Giebels},
  {Glicenstein}, {Gl{\"u}ck}, {G{\"o}ring}, {Grondin}, {H{\"a}ffner}, {Hague},
  {Hahn}, {Hampf}, {Harris}, {Heinz}, {Heinzelmann}, {Henri}, {Hermann},
  {Hillert}, {Hinton}, {Hofmann}, {Hofverberg}, {Holler}, {Horns},
  {Jacholkowska}, {Jahn}, {Jamrozy}, {Jung}, {Kastendieck}, {Katarzy{\'n}ski},
  {Katz}, {Kaufmann}, {Kh{\'e}lifi}, {Klochkov}, {Klu{\'z}niak}, {Kneiske},
  {Komin}, {Kosack}, {Kossakowski}, {Krayzel}, {Laffon}, {Lamanna}, {Lenain},
  {Lennarz}, {Lohse}, {Lopatin}, {Lu}, {Marandon}, {Marcowith}, {Masbou},
  {Maurin}, {Maxted}, {Mayer}, {McComb}, {Medina}, {M{\'e}hault}, {Menzler},
  {Moderski}, {Mohamed}, {Moulin}, {Naumann}, {Naumann-Godo}, {de Naurois},
  {Nedbal}, {Nguyen}, {Niemiec}, {Nolan}, {Ohm}, {de O{\~n}a Wilhelmi},
  {Opitz}, {Ostrowski}, {Oya}, {Panter}, {Parsons}, {Paz Arribas}, {Pekeur},
  {Pelletier}, {Perez}, {Petrucci}, {Peyaud}, {Pita}, {P{\"u}hlhofer}, {Punch},
  {Quirrenbach}, {Raue}, {Reimer}, {Reimer}, {Renaud}, {de los Reyes},
  {Rieger}, {Ripken}, {Rob}, {Rosier-Lees}, {Rowell}, {Rudak}, {Rulten},
  {Sahakian}, {Sanchez}, {Santangelo}, {Schlickeiser}, {Schulz}, {Schwanke},
  {Schwarzburg}, {Schwemmer}, {Sheidaei}, {Skilton}, {Sol}, {Spengler},
  {Stawarz}, {Steenkamp}, {Stegmann}, {Stinzing}, {Stycz}, {Sushch}, {Szostek},
  {Tavernet}, {Terrier}, {Tluczykont}, {Valerius}, {van Eldik}, {Vasileiadis},
  {Venter}, {Viana}, {Vincent}, {V{\"o}lk}, {Volpe}, {Vorobiov}, {Vorster},
  {Wagner}, {Ward}, {White}, {Wierzcholska}, {Wouters}, {Zacharias}, {Zajczyk},
  {Zdziarski}, {Zech}, \& {Zechlin}}]{2013A&A...550A...4H}
{H.E.S.S.~Collaboration}, {Abramowski}, A., {Acero}, F., {et~al.} 2013, \aap,
  550, A4

\bibitem[{{Horns} \& {Meyer}(2012)}]{2012JCAP...02..033H}
{Horns}, D. \& {Meyer}, M. 2012, \jcap, 2, 33

\bibitem[{{Horns} \& {Meyer}(2013)}]{2013arXiv1309.3846H}
{Horns}, D. \& {Meyer}, M. 2013, arXiv:1309.3846

\bibitem[{{Inoue} {et~al.}(2013){Inoue}, {Granot}, {O'Brien}, {Asano},
  {Bouvier}, {Carosi}, {Connaughton}, {Garczarczyk}, {Gilmore}, {Hinton},
  {Inoue}, {Ioka}, {Kakuwa}, {Markoff}, {Murase}, {Osborne}, {Otte},
  {Starling}, {Tajima}, {Teshima}, {Toma}, {Wagner}, {Wijers}, {Williams},
  {Yamamoto}, {Yamazaki}, \& {CTA Consortium}}]{2013APh....43..252I}
{Inoue}, S., {Granot}, J., {O'Brien}, P.~T., {et~al.} 2013, Astroparticle
  Physics, 43, 252

\bibitem[{{Jacob} \& {Piran}(2008)}]{2008PhRvD..78l4010J}
{Jacob}, U. \& {Piran}, T. 2008, \prd, 78, 124010

\bibitem[{{Jauch} \& {Rohrlich}(1976)}]{1976tper.book.....J}
{Jauch}, J.~M. \& {Rohrlich}, F. 1976, {The theory of photons and electrons.
  The relativistic quantum field theory of charged particles with spin
  one-half}

\bibitem[{{Jelley}(1966)}]{1966PhRvL..16..479J}
{Jelley}, J.~V. 1966, Physical Review Letters, 16, 479

\bibitem[{{Kifune}(1999)}]{1999ApJ...518L..21K}
{Kifune}, T. 1999, \apjl, 518, L21

\bibitem[{{Madau} \& {Pozzetti}(2000)}]{2000MNRAS.312L...9M}
{Madau}, P. \& {Pozzetti}, L. 2000, \mnras, 312, L9

\bibitem[{{Mankuzhiyil} {et~al.}(2010){Mankuzhiyil}, {Persic}, \&
  {Tavecchio}}]{2010ApJ...715L..16M}
{Mankuzhiyil}, N., {Persic}, M., \& {Tavecchio}, F. 2010, \apjl, 715, L16

\bibitem[{{Mazin} \& {Raue}(2007)}]{EBLMaRa}
{Mazin}, D. \& {Raue}, M. 2007, \aap, 471, 439

\bibitem[{{Mazin} {et~al.}(2013){Mazin}, {Raue}, {Behera}, {Inoue}, {Inoue},
  {Nakamori}, {Totani}, \& {CTA Consortium}}]{2013APh....43..241M}
{Mazin}, D., {Raue}, M., {Behera}, B., {et~al.} 2013, Astroparticle Physics,
  43, 241

\bibitem[{{M{\'e}sz{\'a}ros}(2013)}]{2013APh....43..134M}
{M{\'e}sz{\'a}ros}, P. 2013, Astroparticle Physics, 43, 134

\bibitem[{{Meyer} {et~al.}(2013){Meyer}, {Horns}, \&
  {Raue}}]{2013PhRvD..87c5027M}
{Meyer}, M., {Horns}, D., \& {Raue}, M. 2013, \prd, 87, 035027

\bibitem[{{Meyer} {et~al.}(2012){Meyer}, {Raue}, {Mazin}, \& {Horns}}]{Meyer12}
{Meyer}, M., {Raue}, M., {Mazin}, D., \& {Horns}, D. 2012, arXiv:1202.2867

\bibitem[{{Neronov} \& {Vovk}(2010)}]{2010Sci...328...73N}
{Neronov}, A. \& {Vovk}, I. 2010, Science, 328, 73

\bibitem[{{Nikishov}(1962)}]{REF::NIKISHOV::JETP1962}
{Nikishov}, A.~I. 1962, Soviet Physics JETP, 14, 393

\bibitem[{{Orr} {et~al.}(2011){Orr}, {Krennrich}, \& {Dwek}}]{Orr}
{Orr}, M.~R., {Krennrich}, F., \& {Dwek}, E. 2011, \apj, 733, 77

\bibitem[{{Prandini} {et~al.}(2010){Prandini}, {Bonnoli}, {Maraschi},
  {Mariotti}, \& {Tavecchio}}]{2010MNRAS.405L..76P}
{Prandini}, E., {Bonnoli}, G., {Maraschi}, L., {Mariotti}, M., \& {Tavecchio},
  F. 2010, \mnras, 405, L76

\bibitem[{{Punch} {et~al.}(1992){Punch}, {Akerlof}, {Cawley}, {Chantell},
  {Fegan}, {Fennell}, {Gaidos}, {Hagan}, {Hillas}, {Jiang}, {Kerrick}, {Lamb},
  {Lawrence}, {Lewis}, {Meyer}, {Mohanty}, {O'Flaherty}, {Reynolds}, {Rovero},
  {Schubnell}, {Sembroski}, {Weekes}, \& {Wilson}}]{Punch92}
{Punch}, M., {Akerlof}, C.~W., {Cawley}, M.~F., {et~al.} 1992, \nat, 358, 477

\bibitem[{{Raue} \& {Meyer}(2012)}]{2012AIPC.1505..610R}
{Raue}, M. \& {Meyer}, M. 2012, in American Institute of Physics Conference
  Series, Vol. 1505, American Institute of Physics Conference Series, ed. F.~A.
  {Aharonian}, W.~{Hofmann}, \& F.~M. {Rieger}, 610--613

\bibitem[{{Reimer}(2007)}]{Reimer2007}
{Reimer}, A. 2007, \apj, 665, 1023

\bibitem[{{Reimer} \& {B{\"o}ttcher}(2013)}]{2013APh....43..103R}
{Reimer}, A. \& {B{\"o}ttcher}, M. 2013, Astroparticle Physics, 43, 103

\bibitem[{{Renault} {et~al.}(2001){Renault}, {Barrau}, {Lagache}, \&
  {Puget}}]{2001A&A...371..771R}
{Renault}, C., {Barrau}, A., {Lagache}, G., \& {Puget}, J.-L. 2001, \aap, 371,
  771

\bibitem[{{Sanchez} {et~al.}(2013){Sanchez}, {Fegan}, \&
  {Giebels}}]{2013A&A...554A..75S}
{Sanchez}, D.~A., {Fegan}, S., \& {Giebels}, B. 2013, \aap, 554, A75

\bibitem[{{S{\'a}nchez-Conde} {et~al.}(2009){S{\'a}nchez-Conde}, {Paneque},
  {Bloom}, {Prada}, \& {Dom{\'{\i}}nguez}}]{2009PhRvD..79l3511S}
{S{\'a}nchez-Conde}, M.~A., {Paneque}, D., {Bloom}, E., {Prada}, F., \&
  {Dom{\'{\i}}nguez}, A. 2009, \prd, 79, 123511

\bibitem[{{Sol} {et~al.}(2013{\natexlab{a}}){Sol}, {Zech}, {Boisson}, {Barres
  de Almeida}, {Biteau}, {Contreras}, {Giebels}, {Hassan}, {Inoue},
  {Katarzy{\'n}ski}, {Krawczynski}, {Mirabal}, {Poutanen}, {Rieger}, {Totani},
  {Benbow}, {Cerruti}, {Errando}, {Fallon}, {de Gouveia Dal Pino}, {Hinton},
  {Inoue}, {Lenain}, {Neronov}, {Takahashi}, {Takami}, {White}, \& {CTA
  Consortium}}]{2013APh....43..215S}
{Sol}, H., {Zech}, A., {Boisson}, C., {et~al.} 2013{\natexlab{a}},
  Astroparticle Physics, 43, 215

\bibitem[{{Sol} {et~al.}(2013{\natexlab{b}}){Sol}, {Zech}, {Boisson},
  {Krawczynski}, {Fallon}, {de Gouveia Dal Pino}, {Hinton}, {Inoue}, {Neronov},
  \& {White}}]{2013IAUS..294..459S}
{Sol}, H., {Zech}, A., {Boisson}, C., {et~al.} 2013{\natexlab{b}}, in IAU
  Symposium, Vol. 294, IAU Symposium, ed. A.~G. {Kosovichev}, E.~{de Gouveia
  Dal Pino}, \& Y.~{Yan}, 459--470

\bibitem[{{Stecker} {et~al.}(1992){Stecker}, {de Jager}, \&
  {Salamon}}]{1992ApJ...390L..49S}
{Stecker}, F.~W., {de Jager}, O.~C., \& {Salamon}, M.~H. 1992, \apjl, 390, L49

\bibitem[{{Stecker} {et~al.}(1993){Stecker}, {de Jager}, \&
  {Salamon}}]{1993AAS...183.2906S}
{Stecker}, F.~W., {de Jager}, O.~C., \& {Salamon}, M.~H. 1993, in Bulletin of
  the American Astronomical Society, Vol.~25, American Astronomical Society
  Meeting Abstracts, 1336

\bibitem[{{Stecker} \& {Glashow}(2001)}]{2001APh....16...97S}
{Stecker}, F.~W. \& {Glashow}, S.~L. 2001, Astroparticle Physics, 16, 97

\bibitem[{{Taylor} {et~al.}(2011){Taylor}, {Vovk}, \&
  {Neronov}}]{2011A&A...529A.144T}
{Taylor}, A.~M., {Vovk}, I., \& {Neronov}, A. 2011, \aap, 529, A144

\bibitem[{{Urry} \& {Padovani}(1995)}]{1995PASP..107..803U}
{Urry}, C.~M. \& {Padovani}, P. 1995, \pasp, 107, 803

\bibitem[{{Vovk} {et~al.}(2012){Vovk}, {Taylor}, {Semikoz}, \&
  {Neronov}}]{2012ApJ...747L..14V}
{Vovk}, I., {Taylor}, A.~M., {Semikoz}, D., \& {Neronov}, A. 2012, \apjl, 747,
  L14

\end{thebibliography}

\end{document}